# Modelling and evaluating travel information during disruptions: An illustrative example from Swedish railways

Abderrahman Ait-Ali [a, *], Anders Peterson [b]

[a] Department of Science and Technology, Linköping University
Luntgatan 2, SE-602 47 Norrköping, Sweden
[b] The Swedish National Road and Transport Research Institute (VTI)
P.O. Box 55685, 114 28 Stockholm, Sweden

[*] Corresponding author. Tel. +46 8 555 367 81; E-mail: abderrahman.ait.ali@liu.se,
ORCID: 0000-0001-9535-0617

**Abstract**

Accurate and timely travel information is an asset for enhancing passenger travel experience during normal traffic, and for mitigating the discomforts during disruptions. With longer and more frequent disruptions as well as increasing ridership, traffic delays can incur substantial costs for passengers and other transport stakeholders, e.g., operators and infrastructure managers. Such costs can, however, be reduced thanks to effective travel information strategies during traffic disruptions. In this paper, we introduce an evaluation model to assess the value of travel information under different scenarios. Focusing on real-time travel information to train passengers, accessibility benefits are quantified in monetary terms based on historical delay distributions, timing of travel information (pre/on-trip) and ridership. Using a case study from the Swedish railways, the model is showcased and applied to a commuter line in Stockholm. The experimental results indicate individual valuations that are higher than references and savings at the system level of at least 23% of the delay costs. Further testing of the model, e.g., on larger-scale scenarios, and including transfer trips, is a possible direction for future works.

## 1. Introduction

With increasing demand for passenger train services, more frequent traffic disruptions and delays result in substantial costs for passengers, operators, and infrastructure managers. Accurate and timely travel information (TI) can, however, reduce these costs by enhancing passenger travel experience and mitigating discomfort, especially during traffic disruptions.

Traffic disruptions in passenger transport systems can be either planned, e.g., for performing maintenance works or special events, or unplanned, e.g., due to accidents or technical failures. Unplanned disruptions are generally more frequent and less predictable, and can have substantial effects on the network, causing delays, cancellations, detours, and dissatisfaction for passengers (Eltved et al., 2021). Providing accurate and timely TI to passengers during these unplanned disruptions can help them adjust their travel plans and reduce the negative impacts of the disruptions (Bruglieri et al., 2015). However, this requires a high level of coordination and communication among different stakeholders, advanced TI systems and effective strategies for the provision of TI.

Earlier studies in cities like Helsinki (Lehtonen and Kulmala, 2002) have shown positive passenger reception of TI systems in passenger and public transport, especially during traffic disruptions, where timely and relevant information proves particularly beneficial to passengers (Cats et al., 2011). Satisfaction with these systems is often monitored by periodically measuring the customer satisfaction index (CSI). Focusing on train passengers in Sweden, a recent survey of CSI shows that passengers state significantly lower satisfaction with TI in the case of disruptions compared to normal traffic (Trafikverket, 2023). The survey results show that the CSI levels during traffic disruptions are significantly lower than the 70% target whereas during normal traffic the levels are around the 80% target. It is therefore important to improve the adopted strategies for the provision of TI to passengers during disruptions. This requires, among others, being able to evaluate the effects and benefits of such TI in the event of unplanned traffic disruptions, leading to passenger delays.

This study uses existing data and valuations to model and evaluate how providing TI to train passengers during unplanned traffic disruptions can affect their travel experience. Previous studies have shown that the availability of TI, such as when passengers receive TI after a disruption occurs, can influence their perceived travel time, satisfaction, and travel choices. This study investigates the impact of different TI availability scenarios on train passengers. The results of this study are relevant for both practice and research. For example, they can help decision-makers decide whether and how to invest in or improve TI systems and strategies. They can also provide a basis for further research on how to optimize the interaction between passengers and TI systems, how to enhance TI strategies, and how to reduce the risks of traffic disruptions.

There are several existing studies on the evaluation of the effects of TI on passengers, with or without focusing on traffic disruptions, see a review by Brakewood and Watkins (2019). Reducing the travel and waiting time (and their uncertainties) as well as increasing CSI are among the main benefits that are emphasized in the literature (Wu et al., 2010, Kattan et al., 2013). However, this requires a high-quality TI system that can provide accurate and timely TI to passengers (Pirra et al., 2017). A commonly adopted approach to evaluate such benefits is using stated preference (SP) surveys, e.g., (Wardman et al., 2001, Dziekan and Vermeulen, 2006, Watkins et al., 2011). Two measures are often adopted to quantify the value of TI to passengers, namely total perceived travel time, e.g., as reduced travel/waiting time (Prather Persson, 1998), and generalized travel cost, e.g., as a monetary value or a percentage of the ticket price (Fearnley et al., 2009).

This study contributes to the existing literature by introducing a novel approach, other than SP surveys, to evaluate TI during disruptions. Focusing on real-time TI to commuter train passengers, various factors are considered in the analysis such as delay distributions (based on historical statistics), timing of TI (travel scenarios), existing valuations (cost parameters) as well as ridership (based on smart card data). Valuations are also presented in monetary terms as accessibility benefits, i.e., reduction in perceived journey times, including travel/waiting times, delays and travel time uncertainty.

The main objective of this study is to model and evaluate the effects and benefits of TI provision to train passengers during traffic disruptions. Based on existing research literature, the impact of different availabilities of real-time TI on passengers is quantified using a monetary evaluation of accessibility benefits. The study explores, therefore, the potential for reducing the delay costs at the system level. Ultimately, the aim is to pave the way for future research and practices to improve TI strategies and mitigate disruption costs.

The literature review revealed the complexity of TI systems. Therefore, several assumptions and simplifications are made leading to some limitations in the study. First, the studied homogeneous cases for TI availability, i.e., pre/on-trip, are a simplification of the multiple scenarios and components of ATIS systems. These simplified cases focus on real-time TI before/after leaving the origin (via station screens/smartphone) which do not capture other TI situations, e.g., other timing, content and or channels. Furthermore, potential variations in passenger characteristics, e.g., socio-economic, and individual factors, are not considered. Such characteristics can influence passenger reactions to TI and therefore on effects and benefits of such information.

Section 2 briefly reviews the existing literature on assessing the effects of TI. The evaluation model is described in section 3. The case study including the results is presented in section 4. Section 5 ends the paper with some concluding remarks including directions for future works.

## 2. Literature Review

The rapid digitalization and development of Information and Communication Technologies (ICTs) has seen notable advancements and applications in transportation sectors, particularly in road, aviation, and to a lesser extent, in railways (Giannopoulos, 2004). The adoption of advanced traveller information systems (ATIS), e.g., used to enhance the quality of car trips (Al-Deek and Kanafani, 1993), is an example of the integration of such ICT tools in transportation. The research literature uses different terms to refer to various types and characteristics of TI, e.g., dynamic, intelligent, advanced or personalized (Chorus, 2012). For the sake of clarity, we use throughout this paper TI to refer to the provided information, and ATIS to refer to the system.

ATIS and the provided TI have different characteristics such as communication channels, information, and recipients. The goal is to assist passengers in moving from their origin to their desired destination in different traffic situations including during disruptions. By providing such guidance, ATIS allows to reduce (disruption-related) travel costs such as travel/waiting time, delays, travel time uncertainty (Müller et al., 2020) and crowding/discomfort (Pritchard, 2018). Assessing such effects and benefits is the subject of several studies in the literature.

To review the existing literature, a systematic methodology was adopted. First, it starts with a comprehensive exploration of the main characteristics of ATIS and the provided TI. Thereafter, the focus is on the effects on passengers, with a specific emphasis on the case of traffic disruptions. The approach involves an examination of relevant studies, primarily focusing on peer-reviewed articles and publications. The selection of references is guided by their relevance to the subject matter, with a preference for recent research and/or studies including quantitative assessments.

This section briefly reviews the main (quality) characteristics of TI with a focus on the case of disruptions. Moreover, it summarises some of the main existing methods and measures to evaluate TI.

### 2.1 Travel information during disruptions

During traffic disruptions, there are several steps to consider when studying the process of TI provision, e.g., information generation, channels, information availability/timing, passengers' reactions and effects (Leng and Corman, 2020). Thus, ATIS as such can be seen as a complex system and the subject of multiple research studies. In their recent review of existing literature, Ait-Ali and Peterson (2023) mention similar important constituents that can define how TI is formed and communicated, namely traffic situation (disruption), information/content, channels, and the users. Existing studies have looked at the characteristics of one or more constituents to investigate how these affect the provision of TI.

The type (and quantity) of information that is provided to passengers during disruptions can directly affect their behaviour and thus lead to different travel effects. Research works have therefore studied what and how much information passengers require in different travel contexts (Tang et al., 2020). In their review of TI behavioural effects, Ben-Elia and Avineri (2015) distinguish between experiential (based on feedback from past experiences), prescriptive (providing guidance), and descriptive information (of travel conditions). An example of the latter is real-time TI which is the focus of several studies in the literature (Dziekan and Kottenhoff, 2007, Cats et al., 2011). Other types of information have also been researched, e.g., real-time crowding information (Pritchard, 2018).

Different ATIS channels are used to convey timely and accurate TI to passengers (Halpern, 2021). These channels include in-vehicle displays/speakers or at terminals/stops/stations, mobile phones or Internet (Zito et al., 2011), and differ in various aspects, e.g., audience, type of messages, cost and portability. Studies based on revealed preferences of passengers while using specific channels, e.g., smartphone applications, examined when and where the channel is used for TI, e.g., Transit App in New York City (Ghahramani and Brakewood, 2016). Others use travellers' stated preferences to find the preferred channels at different stages of the trips, e.g., public transport in Dublin (Caulfield and Mahony, 2007).

Different channels often fall within the responsibility of various stakeholders. This is the case, for instance, in deregulated transport markets such as European railways (Ait-Ali and Eliasson, 2021), where different private companies provide competing train services and therefore use various channels, e.g., SMS, websites/apps. Moreover, the infrastructure manager is often also responsible for TI through other channels such as screens/speakers at large stations (Pirra et al., 2017). Given the multiple existing channels, studies show that the needs and preferences for TI also differ between user groups, e.g., gender, age, socio-economic group (Lappin, 2000, van Essen et al., 2016), passengers with disabilities and elderly (Waara, 2009). In their SP-based study of passengers' TI-seeking behaviour in Great Britain, Yeboah et al. (2019) found that factors such as travel frequency, employment status, mode of transport, and preferred information sources (e.g., Internet) are significant predictors of pre-travel TI seeking. A recent Swedish study by Berggren et al. (2021) shows that pre-trip TI is specifically more important for long trips (> 1 hour) and its use varies by age and gender. The authors also show that trip purpose and time of day partly affect how TI affects travel and waiting times.

During disruptions, effective ATIS provide TI to passengers at different stages of their journey, e.g., before starting the trip, at stations or stops, and while travelling onboard (Titov et al., 2023). Existing research indicates that some of the main quality aspects in the provision of TI include the accuracy, timeliness and comprehensiveness of the provided information at the different stages of the journey (Pirra et al., 2017). Such quality aspects are typically measured using various methods. For instance, passenger surveys are often conducted, also known as SP surveys, to measure how satisfied passengers are with these quality aspects, e.g., CSI levels, see the recent survey by Trafikverket (2023) in Sweden. Other methods are based on the analysis of collected travel data, i.e., revealed preference (RP), or based on simulation models (Leng and Corman, 2020, Müller et al., 2020, Paulsen et al., 2021).

Timeliness is the relevance and usefulness of TI for passengers when they need it, particularly valuable during unplanned traffic disruptions as it enables passengers to make more informed travel decisions (Bruglieri et al., 2015). Timely TI depends on the journey stage, the disruption type, and the TI accuracy. For instance, the latter often has a balance with timeliness, as more timely TI may lead to less accuracy, and vice versa. A recent study by Paulsen et al. (2021) investigated how passengers adapt to different types of timely travel information (TI) during public transport disruptions. They considered various scenarios of TI availability, such as no TI, pre-trip TI, and TI at stops, and measured their effects on passenger delays and route choices in a large-scale, multi-modal transport network of Metropolitan Copenhagen. In a similar study, Leng and Corman (2020) distinguish between other types apart from no TI, namely advanced TI and timely TI. In addition to the case of no TI (never), we focus in this study instead on the cases where TI is made available before or after the trip starts, i.e., leaving for the departure station. The latter two cases are commonly referred to in the literature as pre-trip and on-trip (or en route) information, respectively (Zito et al., 2011, Skoglund, 2012, Berggren et al., 2021).

## 2.2 Assessing the effects of travel information

Besides studies on the characteristics of TI, several research works focused on the effects of such information both on passengers and the network/system as a whole, e.g., see a review by van Essen et al. (2016). Others aimed at assessing these effects or benefits, e.g., see a review by Brakewood and Watkins (2019).

To investigate the impacts of TI on public transport passengers, Skoglund (2012) distinguishes between short-term and long-term effects. The former effects refer to immediate changes in travel behaviour in response to the provided TI, such as departure time or route choice (Wu et al., 2010). On the other hand, long-term effects occur over a longer period and involve more significant behavioural changes, such as travel satisfaction (CSI), and changes in travel demand, e.g., due to modal shifts and/or changes in travel patterns (Kattan et al., 2013). For instance, Zito et al. (2011) find that 15% of public transport passengers in Palermo are willing to travel more if quality TI is available.

An earlier review by van Essen et al. (2016) highlights the importance of considering both the impacts on individual travellers and the system/network effects. The latter can have implications on collective behaviours and transport systems (Ben-Elia and Avineri, 2015). ATIS can therefore be used, especially in the events of traffic disruptions, for network management (Giannopoulos, 2004), e.g., to shift the effects from user equilibrium toward system optimum (van Essen et al., 2016). Although briefly discussed in this study, network and long-term effects are not included in the developed model.

In the case of real-time TI, Brakewood and Watkins (2019) identify two major benefits for individual passengers, namely travel and waiting time. These benefits are the results of several passenger choices resulting from the available information, e.g., choice of travel, mode, route, or departure time. In their study focusing on real-time information at public transport stops, Dziekan and Kottenhoff (2007) identified several major benefits for passengers including reduced waiting time, positive psychological factors related to reduced uncertainty, increased willingness-to-pay (WTP), and higher customer satisfaction (higher CSI). Ferris et al. (2010), Watkins et al. (2011) found similar results based on a survey of smartphone app users in the Seattle area providing real-time arrival and bus location information. The results show benefits such as an increase in overall satisfaction with public transit, decreased waiting time, increased transit trips per week, increased feelings of safety, and even health benefits from increased walking.

To assess the effects and benefits of TI, existing research studies adopt different methods (Lu, 2015). An earlier work by Malchow et al. (1996) attempted to set the analytical basis for studying the economics of TI provision. In a later analytical study, Arentze and Timmermans (2005) measure the expected information gains of individuals when making a trip under conditions of uncertainty such as in the event of traffic disruptions. The authors developed an analytical Bayesian model to measure these gains and used simulations to illustrate and demonstrate their approach. The expected gains are measured as an element of the travel utility function of the different trip choice alternatives. To study the perceived value of TI, Chorus et al. (2006) adopt a similar conceptualization and use simulations with numerical examples for validation.

A common method is the use of SP surveys where passengers are asked about, e.g., perceived waiting time, usage and satisfaction level, or preferences for TI in different (hypothetical) situations (Lu, 2015). To estimate the benefits of TI from a smartphone app for bus trips in the Seattle area, Ferris et al. (2010) use the same method. However, the authors measure the benefits in terms of perceived and actual waiting time. Simulation-based methods are increasingly adopted to assess the benefits of TI, especially in the event of disruptions. Using an agent-based simulation, Leng and Corman (2020) and more recently Paulsen et al. (2021) studied the effects of different TI types on passengers during traffic disruptions. A similar method is adopted by Cats and Gkioulou (2017) to study the impacts of TI (and public transport reliability) on the waiting time of passengers and its uncertainty.

To measure the benefits of ATIS during disruptions, studies use several possible quantitative measures. Travel and waiting time are important components of the total perceived travel time, which is influenced by the availability and quality of TI (Watkins et al., 2011, Berggren et al., 2021). Travel and waiting time variabilities, e.g., due to unplanned delays, are related to the journey time uncertainty and reliability, which can affect travellers' satisfaction and choices (Peer, 2013, Tseng et al., 2013). In the context of appraisals and evaluations, Börjesson and Eliasson (2011) recommend the use of the standard deviation of travel times for the valuation of reliability. Other important components include, for instance, delays, transfer time, and congestion/discomfort. An alternative (but more general) measure is accessibility from a passenger perspective which reflects how easy/hard it is to reach/access a destination (Lu, 2015). A composite measure that is used to capture some accessibility attributes is the total perceived journey time (Prather Persson, 1998, Wardman, 2001). To express these non-monetary attributes, the generalized travel cost is an alternative composite measure in the context of economic appraisals, e.g., cost-benefit analyses (Chorus et al., 2006, Fearnley et al., 2009). WTP is a related monetary measure that is commonly used in the valuation of these different travel attributes (e.g., travel and waiting times). In the case of travel information, several studies have estimated how much travellers are willing to pay for TI in different travel situations. WTP helps thus quantify, in monetary terms, the perceived benefits of information provision (Zito et al., 2011).

## 2.3 Summary

The literature review explored relevant aspects of ATIS including TI characteristics such as channels, information, and users with a focus on passenger transport and quality aspects in the event of disruptions. TI availability, which is particularly important in the event of disruptions, is often measured through surveys of TI comprehensiveness and passenger satisfaction and to a lesser extent using simulation models. See Table 1 for an overview of selected references. These are mainly the studies which include quantitative assessment, using a certain measure, of the value of TI availability to passengers.

ATIS and TI have both short-term and long-term effects. It is also important to consider both individual and system effects, e.g., for network management in the event of traffic disruptions. The provision of real-time TI has major benefits for passengers, these include reduced travel and waiting time as well as increased satisfaction with the transport system.

The review also covers several methods to assess the effects and benefits of TI provision during disruptions. SP surveys are commonly used in the literature. Unlike RP surveys which analyse actual behaviours, SP surveys capture individual preferences that are stated by passengers. Simulation-based methods are increasingly adopted as they allow for the creation of controlled virtual scenarios. The existing literature also includes different measures to assess the value of TI for passengers, such as travel and waiting time, total perceived journey time, generalized travel costs, and WTP.

Selected studies from the literature, that are included in Table 1, all have explicit valuations of TI using different measures. These studies indicate that the WTP for TI lies between 5 and 20% of the ticket price per trip. Alternative measures are also used such as accessibility-based quantities, e.g., travel and waiting time, and satisfaction levels (CSI). Moreover, these studies focus on real-time information for public transport passengers in different countries. Until recently, SP surveys using questionnaires have been the predominant approach to do the assessments. Newer approaches include the use of simulations together with SP surveys, and smartphone apps. Few studies have studied the case of traffic disruptions, e.g., (Cats and Gkioulou, 2017).

As Table 1 shows, most reviewed studies in the literature focus on real-time TI, their general comprehensiveness and passenger satisfaction (CSI) across different countries and passenger transport systems. SP surveys with questionnaires seem to be the main approach, to assess passengers' preferences and WTP. Conducting such surveys can, however, be costly, especially for large passenger transport systems. Some recent studies have also employed simulations and smartphone apps as a supplement, but few studies have focused on the case of unplanned traffic disruptions, which are shown to greatly influence passengers' travel experience and behaviour (Cats and Gkioulou, 2017). Moreover, the economic impacts, e.g., using cost-benefit analysis, of ATIS have been scarcely explored. Considering the growing availability of data from new technologies, e.g., offering valuable information on passenger and vehicle movements, there is a need for novel alternative approaches to assessing TI using existing data.

Table 1. Overview of selected studies showing the studied TI, the adopted assessment method and explicit valuation.

| Reference (Chronologically) | Studied TI (What?) | Assessment method (How?) | Valuation (How much?) |
|---|---|---|---|
| (Prather Persson, 1998) | Real-time at train stations in Sweden | SP-survey (questionnaires) | 3 min of travel time (speakers) 6 min (screens) |
| (Wardman et al., 2001) | Real-time at interchange terminals in the UK | SP-survey (questionnaires) | 1.4 min of the in-vehicle travel time (or about 5 British Pence per trip) |
| (Dziekan and Vermeulen, 2006) | Real-time displays at tram stations in the Hague | SP-survey (questionnaires before and after implementation) | 20% of the perceived waiting time (from 6.3 to 5 min, or 1.3 min) |
| (Fearnley et al., 2009) | Real-time system for public transport in Oslo | SP-survey (questionnaires) | 2.37 Norwegian crown per trip |
| (Watkins et al., 2011) | Real-time on the smartphone for buses in Seattle | SP survey (using a smartphone app) | 30% of the average waiting time (from 9.9 to 7.5 min, or 2.4 min) |
| (Zito et al., 2011) | Pre-trip and on-trip real-time | SP survey (and Monte Carlo simulations for uncertainty analysis) | 15% of passengers are willing to increase their use of public transport |
| (Brakewood et al., 2014) | Real-time via web/mobile phones in Tampa | SP survey (using a smartphone app and web-based surveys) | 2 min reduction in waiting time |
| (Cats and Gkioulou, 2017) | Real-time in Stockholm's public transport network under disruptions | Agent-based simulations | 14% of the average perceived waiting time |

The current study fills, therefore, several gaps in the existing literature. First, the focus is on the valuation during unplanned traffic disruptions. Moreover, we introduce a new approach combining historical traffic and passenger data with existing valuations from previous SP surveys, e.g., on the value of travel and waiting time, delays and travel uncertainty. For more practical and system comparison, e.g., in cost-benefit analyses of ATIS investments and TI strategies, the valuations are provided explicitly using accessibility-based monetary measures.

## 3. Modelling and evaluating travel information

The adopted modelling framework, for evaluating the effect of TI during disruptions, includes modelling of the problem, the effect of TI, and monetary measures. The goal of the problem modelling is to describe the traffic characteristics by reading relevant data including timetable, ridership data, and disruption distributions, i.e., delay length and frequencies affecting passengers. TI effects are modelled as an increase in passenger accessibility, i.e., reduced perceived journey time, and hence monetary benefits of different availabilities of TI. These effects are quantified and aggregated (with/without ridership data) in monetary values from passenger generalized travel costs.

What follows is a more detailed presentation of the modelling framework as well as discussions of some of the main assumptions.

### 3.1 Problem description

There are several possible travel scenarios for passengers during traffic disruptions affecting passengers willing to travel from an origin to a destination, e.g., for work, shopping, or any other leisure activity. Moreover, the passengers' trip includes several steps, such as walking from the origin (e.g., home, work, or other activity location) to the starting station, waiting for departure, in-vehicle travel (onboard) to the destination station and a final walk to the destination (work, or home, or other activity location). Using a space-time diagram, Figure 1 presents an illustrative example where the vertical or space axis includes some of the trip steps on the original and delayed routes, e.g., walking and travel onboard.

For the sake of simplicity, we will consider the original route as the best path when there are no traffic disruptions. This route is considered best from the traveller's perspective, i.e., minimizes the total journey time. In case of delays on this route, passengers can make changes to their original travel plan depending on when TI is made available and/or the delay length. For instance, Figure 1 shows how passengers can adapt their original route plan and choose one of the delayed routes, e.g., by either waiting at home or the departure station depending on the available TI, if any.

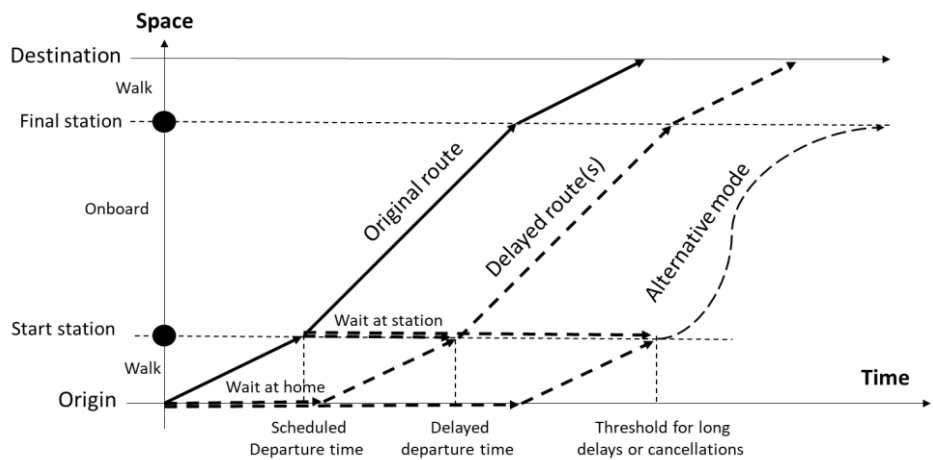

Figure 1. Problem description showing the original and delayed route(s) as well as the routes using an alternative mode.

Note that in this simplification, we only consider direct trips and assess TI on a commuting line as in the case study. However, transfers and connecting trips can also be considered to study larger networks. Moreover, passengers are assumed to choose, if disruptions are severe enough, i.e., longer delays or cancelled departure, to travel using other modes of transportation, e.g., replacement buses. Based on some existing studies, e.g., (Teng et al., 2018), a threshold waiting time of 30 minutes is reasonable in practice. However, note that the validity of this threshold (for disruption severity and replacement modes) may vary across space and time, depending on the available alternative routes. For example, passengers from central stations and/or during peak hours may find a 15-minute delay a severe disruption since they have more available travel options.

As mentioned earlier, the passenger's choice of route plan may depend on factors such as the disruption characteristics and the TI availability. In this framework, we focus on the delay distribution, i.e., the length and frequency/probabilities of specific delay lengths based on historical delay statistics. Moreover, we consider TI availability as the timing in the journey when the information is provided and made available to the passenger. In this framework, the following three cases are studied: 1) never, i.e., when no information was ever made available, 2) on-trip TI after leaving the origin and starting the trip, e.g., after leaving home for the departure station, and 3) pre-trip information before trip starts, e.g., the passenger is still at home. The first case, which is rare in practice, is used as a reference for passenger valuation of TI availability in the other two cases. Table 2 shows these cases with examples of TI and the assumed route plan for each.

Based on the table, passengers in the studied problem are assumed to receive a perfect TI, i.e., informed with the correct predictions of, e.g., delay lengths for the original and availability of the alternative routes. Based on this information, the corresponding route plan is assumed to minimize their total journey time. However, other possible scenarios and assumptions can be studied using the same framework. For instance, different variants of TI can be considered, such as incorrect or partially correct predictions. Moreover, specific TI strategies to nudge passengers to adopt a certain route can also be investigated, e.g., to mitigate congestion during disruptions. Thus, the studied cases and assumptions in Table 2 are for the sake of illustration and simplicity, and other types of TI and assumptions can be studied.

Table 2. Studied cases of information availability with examples of TI, and the corresponding assumed route plan.

| Case | Information availability (examples) | The corresponding assumed route plan |
| --- | --- | --- |
| **Never** | No information is available | Passengers continue with the main route and wait for the delayed departure *at the start station* with *uncertain* travel time. |
| **On-trip** | Real-time information (on displays/speakers at stations) | Passengers may continue with the main route and wait for the delayed departure *at the start station*. |
| **Pre-trip** | Real-time information (on website/mobile phones) | Passengers may continue with the main route and wait for the delayed departure *at the origin*. |

Although not included in Table 2, passengers may consider an alternative mode of transportation such as (replacement) buses. For instance, this is the case when disruptions are severe as well as if the alternative mode allows one to reach the final station earlier (than taking a delayed route), see Figure 1 for an illustration. In the pre-trip case, passengers get TI before they leave their origin, i.e., no later than scheduled departure minus walk time to the start station, it is otherwise considered an on-trip TI availability.

### 3.2 Accessibility-based monetary measure

Two composite measures are commonly used in the literature for quantifying the value of TI, namely: total perceived travel time, and generalized travel costs. The former expresses the valuations in terms of improved accessibility, i.e., saved perceived travel time, whereas the latter is a monetary conversion using economic valuations.

Adopting similar accessibility measures, we include the following terms in the total perceived travel time for a passenger who is affected by the disruptions and/or TI availability:

- Waiting time $w$ can be collected from average headway using the timetable. For frequent services and/or unavailable TI (never), it can be assumed to be half of the headway.
- Scheduled in-vehicle travel time $t$ can also be collected using timetable data.
- Delay length $d$ can be calculated using data on historical delay statistics.
- Travel time extension $\gamma t_{sd}$ because of uncertainty (due to the lack of TI) where $\gamma < 1$ is a factor and $t_{sd}$ is the standard deviation in travel times due to traffic disruptions.

Using these terms, the total perceived travel time can be formulated as in equation (1). However, other terms may also be included in the formulation such as the crowding factor to capture the prolonged perceived in-vehicle time because of increased travel discomfort from high numbers of passengers onboard.

$$A = w + d + t + \gamma t_{sd} \qquad (1)$$

Similarly, the monetary valuation in terms of the generalized travel cost is calculated using the corresponding cost parameters, i.e., $\alpha_t$ as the value of in-vehicle travel time, $\alpha_w$ and $\alpha'_w$ as the value of passive and active waiting time, respectively. Thus, the monetary valuation in this framework uses the formulation in equation (2). The formulation illustrates the generalized travel cost for a passenger who knows the departure schedule but does not know about the departure delay, i.e., TI is not available.

$$C = \alpha_w w + \alpha'_w d + \alpha_t (t + \gamma t_{sd}) \qquad (2)$$

In the delayed route plans, passengers can either wait at home (passive waiting time) or at the station (active or transfer waiting time) depending on TI availability. Empirical data shows that the corresponding parameter may differ significantly in value, especially when the frequency is sufficiently low, e.g., less than 5 departures per hour, and when passengers can accurately predict the departure time (Asplund, 2021), e.g., available pre-trip TI. Therefore, we distinguish in this study between the two values of waiting time, i.e., $\alpha_w$ and $\alpha'_w$.

National guidelines exist to translate travel accessibility parameters into monetary values. Many valuations from these guidelines are mainly based on econometric studies of SP or RP surveys. For instance, the evaluation guidelines by the Swedish Transport Administration, Trafikverket (2020), consider $\gamma$ is typically 90% of the value of travel time $\alpha_t$. A summary of some of the adopted values is presented in Table 3 (2017-price level). These values are based on the guidelines by Trafikverket (2020) with a focus on commuter rail traffic. Note these valuations are an average of the values for different trip purposes, e.g., leisure, work, and other purposes.

Table 3. Adopted cost parameters from evaluation guidelines by Trafikverket (2020).

| Cost parameter | Notation | Adopted value in SEK[1] per person-hour |
|---|---|---|
| In-vehicle travel time | $\alpha_t$ | 71 |
| Passive waiting time | $\alpha_w$ | 81.5 for $w \leq 10$ min |
| | | 66.5 for $w > 10$ min |
| Active waiting time | $\alpha'_w$ | 178.5 |
| Uncertainty (of travel time) | $\gamma$ | 0.9 (no unit) |

[1] 1 Euro is around 11 SEK or Swedish crowns (kr).

Note that other cost terms can be included in the generalized costs such as ticket prices, walking time from/to origin/destination and direct delay costs. However, these costs cancel out when comparing the studied cases since they are assumed not to vary between the cases and/or do not depend on information availability nor the disruptions delay.

### 3.3 Modelling the effects of travel information

Let $C_i(d)$ be the generalized travel cost for a passenger during traffic disruption length $d$, with TI availability $i \in I$, where $I$ is the set of the studied information availability cases, namely never ($i = 0$), on-trip ($i = 1$), and pre-trip ($i = 2$).

Based on the different assumed route choices (see Table 2), the generalized costs, for different TI availabilities, can be summarized and presented in Table 4 for a given delay length $d$. For instance, in the case of travellers with available pre-trip information, the waiting time at the station $w$ can be assumed to be zero.

Table 4. Generalized costs for different TI availabilities and delay lengths $d$.

| TI availability $i$ | Generalized cost of the changed main route, $C_i(d)$ |
| --- | --- |
| 0. Never | $C_0(d) = \alpha_w w + \alpha'_w d + \alpha_t(t + \gamma t_{std})$ |
| 1. On-trip | $C_1(d) = \alpha_w w + \alpha'_w d + \alpha_t t$ |
| 2. Pre-trip | $C_2(d) = \alpha_w(w + d) + \alpha_t t$ |

The resulting formulation assumes specific types and behaviour of travellers in different scenarios. In real cases, other types/behaviour of travellers may exist such as those who are partially informed or well-informed but behave differently.

A similar table can also provide the relations in terms of accessibility measure, i.e., without multiplying with the cost parameters. As noted earlier, in case of longer delays (or cancellations), the passenger may travel using the best available alternative mode of transportation, e.g., replacement buses. Thus, the in-vehicle travel time $t$ and hence the corresponding generalized cost will depend on other characteristics such as average speed and travelled distance of the alternative mode.

To simplify the problem setup (from Figure 1), passengers can either choose the original route or else another alternative, i.e., either delayed route or other alternative mode of transportation in case of severe disruptions. The simplified setup is represented in Figure 2 where $p = p(d)$ is the probability of having a certain disruption affecting the departure on the original route with a delay length $d$.

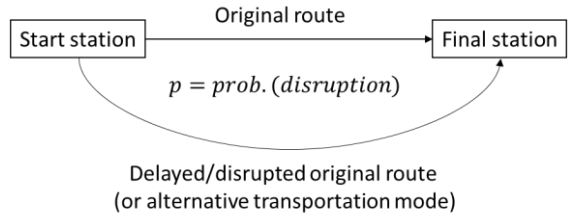

Figure 2. Simplified illustration of the modelled setup.

Let $\Delta C_i = \Delta C_i(d) = C_0(d) - C_i(d)$ represent the gain in generalized cost (or alternatively $\Delta A_i = A_0 - A_i$ for accessibility measures) when TI is available, i.e., $i > 0$, compared to when it is not, i.e., $i = 0$. It has been shown earlier in the literature, e.g., by Malchow et al.

(1996), that the TI value $VT_i$ can be written, as the probability of delay $p$ multiplied by the benefits/gains $\Delta C_i$ from taking the better route. Thus, for a passenger travelling on a route that is affected by disruption with distribution $p$, the value of TI can be formulated as in equation (3).

$$VT_i = \int p \Delta C_i; \quad \forall i > 0 \tag{3}$$

The previous formulation of $VT$ is for a single passenger/trip from an origin to a destination. The total aggregated value of TI should include the aggregation of values for all the travellers and/or all the origin-destination (OD) pairs on the studied line or network. For this, let $VT_i^{(k)}$ denote the value for a passenger travelling the OD pair $k = (o, d)$ where $o$ and $d$ are, respectively, the start and final station. The aggregated value $VT_i^{agg}$ can be calculated using the average over all the line the studied line, i.e., over all the OD-pairs on the line. Moreover, if ridership or OD data is available, e.g., using smart card data for commuter traffic, the average can be weighted, and the aggregated value is formulated as in equation (4), where $n_{o,d}$ is the number of passengers travelling over the OD pair.

$$VT_i^{agg} = \frac{\sum_{k=(o,d)} n_{o,d} \cdot VT_i^{(k)}}{\sum_{o,d} n_{o,d}} \tag{4}$$

Note that the formulated valuation $VT_i^{agg}$ here provides the average value per trip/passenger based on the accumulated value over a specific space and time, e.g., line with/out ridership data on OD-pairs during a typical weekday.

## 4. Illustrative example: Stockholm's commuter rail

In this section, we present some experimental results for the valuation of TI during disruptions.

The commuter rail system in Stockholm during 2015, see the map in Figure 3, is chosen as the case study. Due to limited data availability, the focus is on the line between Nynäshamn (Nyh) in the southeast and Bålsta (Bål) in the northwest. This line consists of 27 stations, and both terminal stations are highlighted in the figure.

The experiments in this study provide two different perspectives. The individual perspective focuses on the individual valuation of TI whereas the system perspective shows how these valuations compare to the total costs of traffic disruptions in the system. Both perspectives are studied based on the studied line from Stockholm's commuter network as well as during the year 2015.

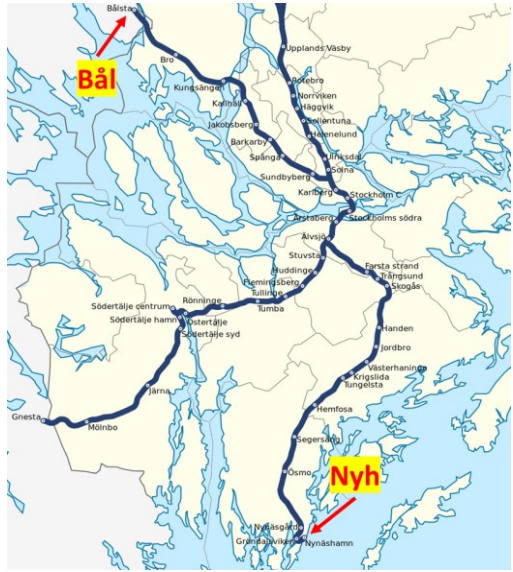

Figure 3. Map of the 2015 commuter rail in Stockholm showing the studied line between Bål (in the northwest) and Nyh in the south (Frohne et al., 2014).

## 4.1 Input data

Data for this study was collected from various sources. For example, values for the cost parameters are based on the guidelines by Trafikverket (2020), see Table 3 for the adopted cost parameter values.

Based on the 2015 winter timetable, the average waiting time on a typical weekday is around 15 minutes for all stations between Kän (Kungsängen) and Vhe (Västerhaninge), and around 22.5 minutes for all the other stations on the studied line. The average travel time between two consecutive stations, which varies between 2 and 6 minutes, as well as the dwell time are also collected from the same timetable, see the heatmap to the left in Figure 4 for the total travel time (in minutes) between pairs of stations on the studied line.

In case of severe disruption, i.e., cancellation or delays longer than 30 min, we assume that passengers choose an alternative transportation mode, i.e., replacement buses, as the best alternative route. Travel times for this route are based on the average bus speed and the distances between stations. Such average speed is assumed to be 40 km/h, which is also the average speed of public transport services in Sweden (Kenworthy, 2020). These distances are collected using the Google Maps API and are illustrated in the heatmap to the right in Figure 4 in kilometres. The contours in the figure correspond to the pairs with similar values.

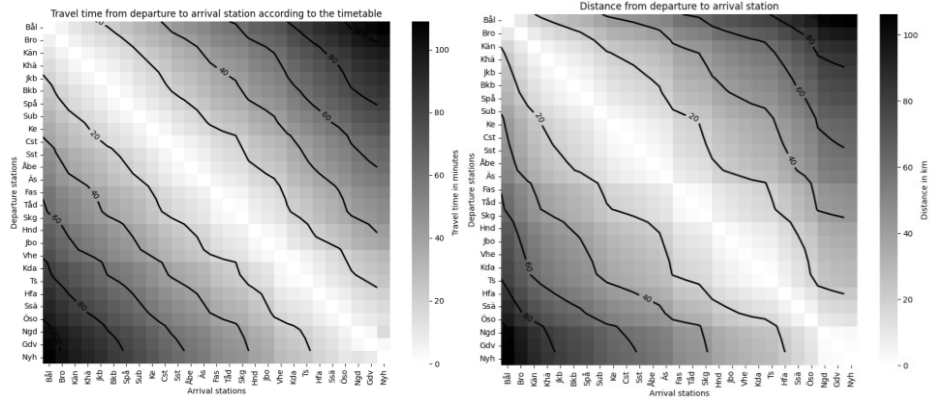

Figure 4. (To the left) Travel time from train timetable. (To the right) road distance for buses between the stations on the studied line.

Traffic data during 2015, providing delay statistics based on around 758 220 train movement observations, is used to estimate the average departure delays and their probability distribution for all the studied stations and both train directions (north/southbound). except for terminal stations. Descriptive statistics are summarized in Figure 5. The left figure presents the average departure delays per station and direction whereas the right figure similarly presents the standard deviations. Note that Bål (Bålsta) and Nyh (Nynäshamn), which are terminal stations on the studied line, have missing values for northbound and southbound directions, respectively.

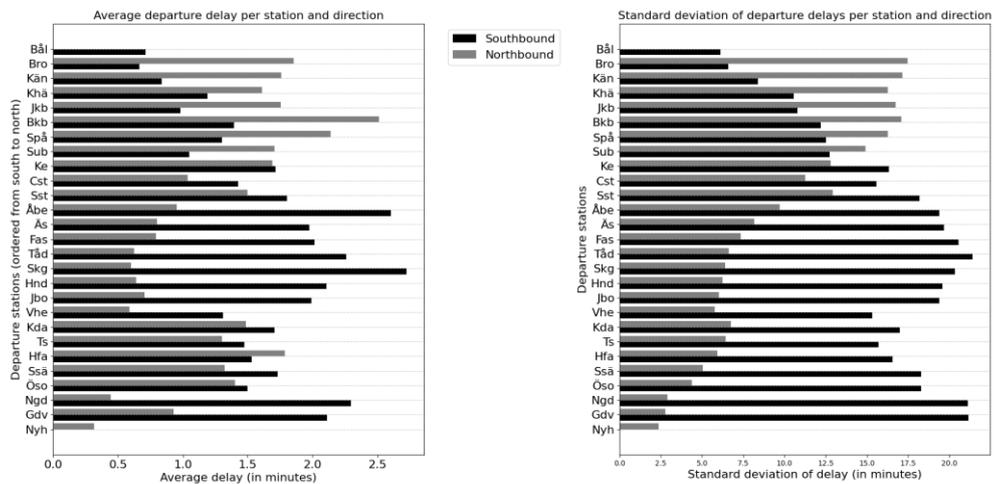

Figure 5. Average and standard deviation of departure delays per station and direction.

Based on Figure 5, longer delays occur later in the trips, e.g., average delays for the southbound direction are lower closer to the first station and vice versa. Additionally, the variance of the delays, which is important for the cost of travel time uncertainty, decreases closer to the first departure station. A more detailed description of the delay is presented in Figure 6. The figure illustrates the delay distribution function per direction for three selected stations, i.e., Cst (central station with the highest number of boarding and alighting passengers), Bkb and Skg which have, respectively, the highest average delay for northbound and southbound directions. Note that all earlier departures are assumed to be on time, i.e., no negative delay. Additionally, long delays (> 30 min) and departure cancellations are all considered as 30 min late, see the sudden increase in Figure 6 at 30 min departure delay.

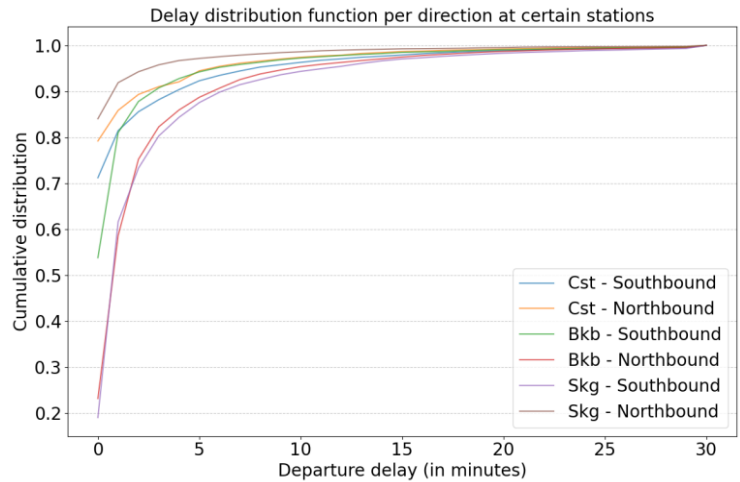

Figure 6. Delay distribution at Stockholm central station, based on 2015 traffic data.

Another type of data, which is used in this case study, is the estimated daily ridership counts, i.e., OD between station pairs on the studied line. Such estimates are based on information from automatic fare collection (AFC) systems, e.g., smart cards. For more details, interested readers are referred to a study by Ait-Ali and Eliasson (2019) who use boarding AFC data to estimate such OD data for Stockholm's commuter rail network. Figure 7 shows the daily ridership from and to the different stations on the studied line. Over a million daily passengers travel between the stations, and the highest number of passengers depart/arrive from/to Cst.

Note that we assume that no trips exist from and to the same station, hence there is no value for TI on these pairs. Moreover, this ridership data is mainly used in the aggregation of the value of TI. However, such data can also be used to estimate the increase in travel discomfort if considered in the model.

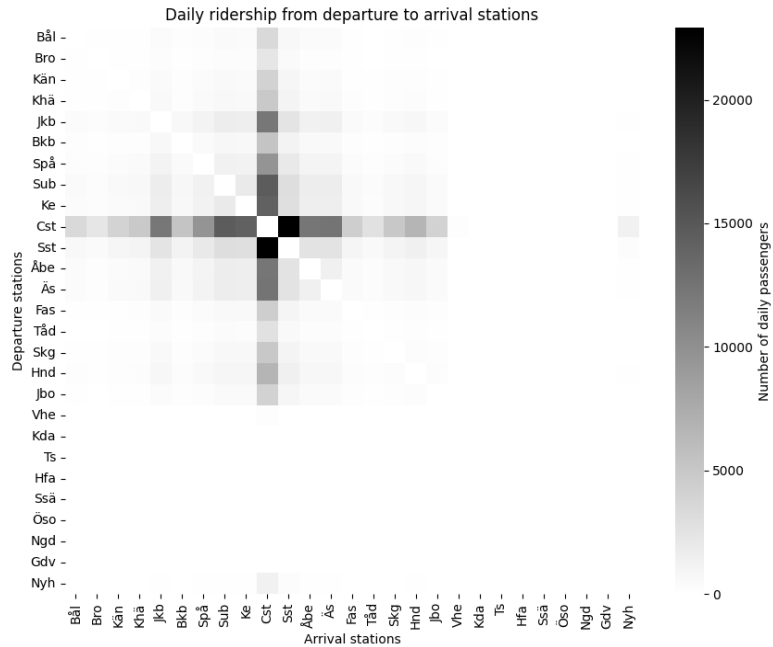

Figure 7. Daily ridership from and to the studied stations, i.e., OD data, during a typical weekday in the 2015 winter period.

## 4.2 Individual perspective: valuation of TI availability during disruptions

Using the collected input data, the individual perspective of the case study focuses on the resulting values of TI during disruptions per individual passengers/trips. An illustration of such results is provided in Figure 8 for both pre-trip (in dark colour) and on-trip TI (in grey colour). For the different departure stations, the figure presents the average benefits of TI availability. The benefits are expressed in SEK per started trip from the corresponding departure station. Note that the departure stations are sorted in descending order of benefits, from left to right, e.g., the terminal stations (Nyh and Bål) have the lowest average benefits as expected.

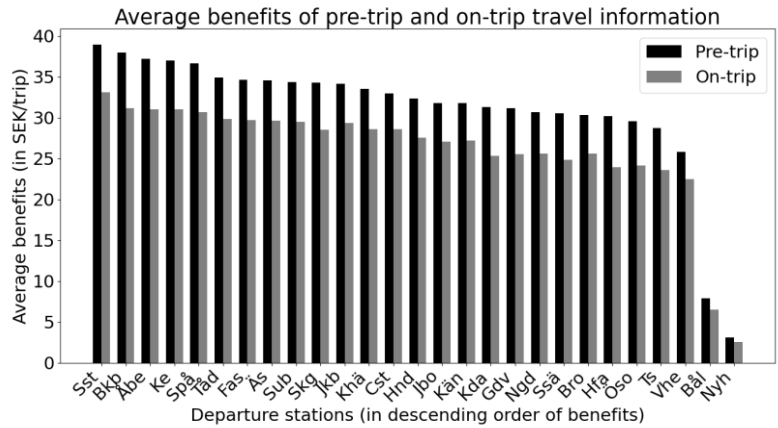

Figure 8. Average benefits of pre-trip and on-trip TI for the different departure stations on the studied line.

The average benefits of TI availability vary between 3 and 40 SEK per trip. As expected, the values are higher for pre-trip information compared to on-trip. Moreover, these average benefits are calculated without accounting for the ridership data, e.g., the number of passengers boarding/alighting. This explains why the central station is not among the departure stations where passengers get the most benefits out of available TI, e.g., Sst and Bkb.

To account for the existing variations in passenger ridership between the different stations, we use an OD-weighted average where the average benefits per departure station are weighted using the number of passengers travelling between that station and other stations on the studied line. The OD-weighted average benefits are presented in Figure 9. As in the previous figure, the departure stations are sorted in descending order of average benefits (of pre-trip TI). Note, however, that these weighted-average benefits are generally lower than the unweighted ones (in Figure 8) since the distribution of alighting passengers, in the ridership data, is not uniform. This can also be noticed, later in this section, when calculating the aggregated value over the studied line.

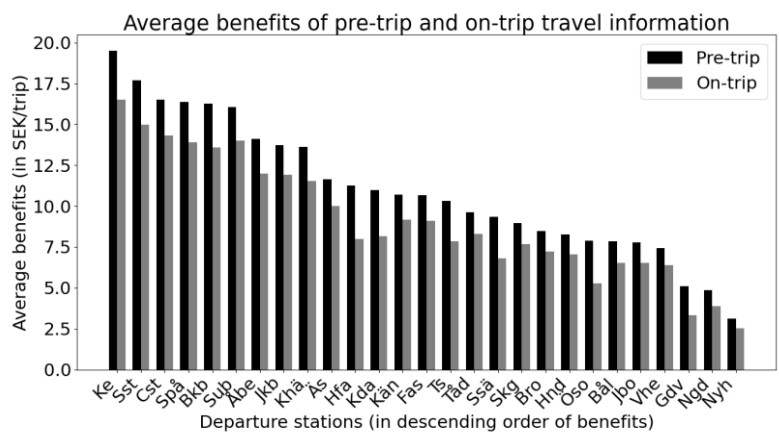

Figure 9. The OD-weighted average benefit of TI availability.

An aggregated value (noted earlier $VT_i^{agg}$) is calculated by considering the average value over all the departure stations. Table 5 shows the resulting aggregated values for on-trip and pre-trip TI both with and without the use of weights from ridership data.

Table 5. Aggregated value $VT_i^{agg}$ (in SEK/trip) of TI.

| Information availability | Weighted average | |
|---|---|---|
| | No | Yes |
| On-trip | 13.0 | 12.7 |
| Pre-trip | 15.5 | 14.8 |

The aggregated values of TI on the studied line are between 12.7 and 15.5 SEK per trip. On-trip TI has a slightly lower value (around 2-2.5 SEK lower) than pre-trip information. This is mainly because of the active waiting time which can be avoided (or made less costly) when pre-trip TI is available. Additionally, the aggregation using ridership data leads to lower values since the values per departure station are weighted depending on the daily number of boarding passengers. The latter, as for alighting passengers, is not uniformly distributed across the studied stations, see Figure 7.

Compared to the references in the literature review (see Table 1), the values in Table 5 correspond to around 4.2-5.2 minutes of (active) waiting time, or around 10-13 minutes of in-vehicle travel time, which is almost twice as high as the reference values in the reviewed literature (Prather Persson, 1998)(Dziekan and Vermeulen, 2006, Watkins et al., 2011). This may be explained by the specificities of the chosen case study, e.g., the studied line, ridership, year of traffic and the distribution of delays, and/or the differences in the evaluation approach that is used in the literature. Another possible explanation is that the passengers' value of travel/waiting time has increased over the years, due to (but not limited to) factors such as increasing travel distances, congestion and income (Rich and Vandet, 2019).

### 4.3 System perspective: comparing TI benefits with delay costs.

To put the calculated benefits of TI availability into a system perspective, we illustrate how these valuations can be used to make informed decisions about TI strategies from a system point of view.

First, we use ridership data to calculate the total daily benefits of TI availability, per departure station, for all boarding and alighting passengers during the studied typical weekday. For each departure station, the resulting daily benefits are presented in Figure 10 in million SEK per weekday.

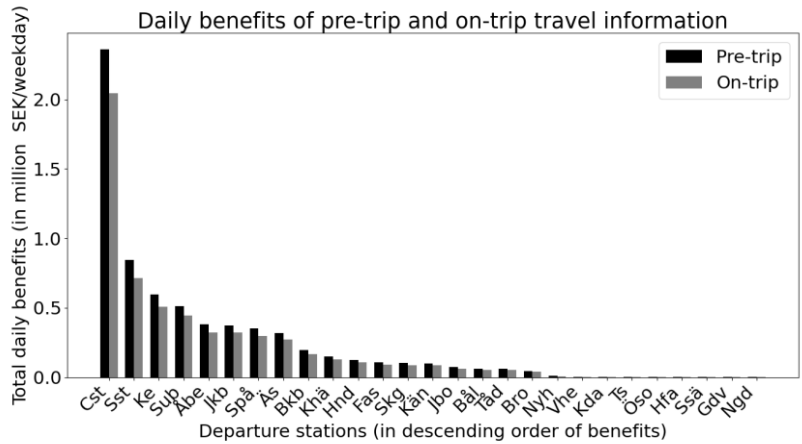

Figure 10. Daily benefits from pre-trip and on-trip information per departure station considering daily ridership.

The departure stations in Figure 10 are presented in order of total daily benefits from the highest (to the left). As expected, the highest benefits are from available pre-trip information at the central station (Cst), with a substantial share of the total benefits, i.e., around 2.4 million SEK (or 28% of all the daily benefits on the studied line). This explains the importance of providing TI to passengers departing from stations with high ridership and/or disruptions. Moreover, the difference, in daily benefits, between pre-trip and on-trip TI is further amplified at the central station. Based on the previous results, it is possible to make more cost-effective prioritisation choices for where and what TI to deploy on a specific line or the passenger transport system in general.

From the individual perspective, the aggregation focused on the un/weighted-average value for both directions. However, it is possible to calculate valuations of TI availability that are specific to the direction of travel. These can be useful to adapt TI strategies to the travel directions in the studied commuter transport system. With a focus on the studied line, Table 6 presents these valuations, in SEK/trip, per direction (southbound or northbound) with/without weighting the average using ridership data. The table results are more detailed compared to the aggregated values in Table 5 which can be obtained by taking the simple or weighted average of the two travel directions.

Table 6. The aggregated value of TI per travel direction (in SEK/trip).

| | | Weighted average | |
|---|---|---|---|
| **Information availability** | **Travel direction** | **No** | **Yes** |
| On-trip | Northbound | 9.7 | 10.8 |
| | Southbound | 16.3 | 14.6 |
| Pre-trip | Northbound | 11.8 | 12.7 |
| | Southbound | 19.1 | 16.9 |

The table results indicate that TI availability has more value in the southbound direction, whether the average is weighted or not. This can be, partly, explained by the fact that the northbound direction has fewer disruptions, see descriptive statics of delays for both directions in Figure 5. The results also indicate that weighting (using ridership data) tends to slightly centralise the averages, i.e., increase lower averages and decrease higher ones, unlike Table 5 where it slightly reduces the values.

Using the national guidelines by Trafikverket (2020), it is possible to calculate the total costs of traffic delays on the studied line and compare them to the benefits of TI availability. The guidelines recommend the use of an average delay cost parameter of 350% of the value of in-vehicle travel time previously noted $\alpha_t$, around 248,5 SEK per passenger-hour of delay. Assuming an average of 255 weekdays in 2015, and adopting a similar calculation methodology as Ahlberg (2015), Table 7 presents the total direct costs of traffic delay on passengers. The table compares these costs to the benefits from on-trip and pre-trip TI per weekday.

Table 7. Average direct delay costs compared to ATIS benefits during a weekday.

|  | Delay costs | Benefits from TI availability | |
|---|---|---|---|
|  |  | On-trip information | Pre-trip information |
| Total in million SEK/weekday | 24.9 | 5.8 (23.3% of delay costs) | 6.8 (27.2% of delay costs) |

The results in the table indicate that an ATIS system, providing pre/on-trip TI to passengers, could have reduced the average direct costs of delays (in 2015) by at least 23% on the studied line. Additional savings can be expected if transfers are also considered, e.g., when studying multiple connected lines in the commuter network.

A more comprehensive cost-benefit analysis of the ATIS system would involve, among others, a comparison of the lifecycle costs (LCC) of the system with its benefits to the overall passenger transport systems over its lifetime. Although a simpler analysis is done here, due to the lack of data and more research on the wider spatiotemporal impacts of TI provision to passengers, the study results provide decision-making guidance and several policy recommendations. First, the economic viability of an ATIS system varies depending on, among others, the network characteristics, service reliability, and the availability of alternative modes and routes. Therefore, a more generalizable conclusion requires a detailed case study including these variables. Nevertheless, based on our results from a specific case study, we estimate that the ATIS system would mitigate reliability costs by around 25%. Assuming that the capital and operating costs of such a system are lower than these costs, it would be worth the investment.

Second and based on the findings of the study, TI value seems to be higher where passenger ridership is high, and where delays are larger and/or less predictable (higher variance). This is because it implies higher total reliability costs for the passengers. TI value may therefore be affected by the magnitude and predictability of passenger delays. and the overall ATIS benefits would depend, among others, on the time of the day (peak/off-peak), location (stops/stations), weather conditions, and passengers' travel preferences. Thus, the need for a more refined and disaggregate analysis accounting for the heterogeneity of passengers and travel scenarios. The results of our (aggregate) analyses suggest, however, that ATIS should prioritize providing accurate and timely information where, in space and time, it would have a larger impact on the total reliability costs.

A closely related implication is that TI to passengers can be seen as an alternative or complementary approach to improve the reliability of the transport systems. The costs of ATIS should be compared to other infrastructure or operational measures to reduce delays. Although it increases passenger travel time costs, adding buffer times to the timetable is an operational measure that would reduce the variability/unpredictability and hence reliability costs. Another example is increased investment in (preventive) maintenance of the infrastructure and vehicles. It would therefore be of interest to study the trade-offs between these measures and the design and implementation of ATIS strategies.

# 5. Concluding remarks

In this concluding section, some of the main highlights of the study are summarized. Limitations and examples of possible future works are also presented.

## 5.1 Highlights

The review of the literature highlights the multiple components and complexity of ATIS systems, particularly during traffic disruptions. Several studies have focused on real-time information to passengers and their subsequent effects, including reduced travel and waiting time, and altered travel behaviour. To evaluate these effects, methodological approaches based on stated or revealed preference surveys are dominant with limited focus on the case of traffic disruptions. In this context and with a focus on traffic delays, this paper introduces an evaluation model that combines historical data and existing valuation results.

Based on the input data (timetable, delay, and ridership), the presented model evaluates the main effects of pre-trip and on-trip TI on passengers during traffic delays. The evaluation of the effects accounts for changes in waiting time, perceived in-vehicle travel time, delay distributions (length and frequency) as well as the uncertainty in travel time due to unavailable TI. Using valuation parameters from the national guidelines, the model output corresponds to the monetary benefits of improved travel accessibility.

A case study from 2015 Stockholm's commuter rail is chosen to test the model and illustrate results from both individual and system perspectives. Focusing on a specific line of the commuter network and using available data, the individual perspective assesses the value of pre-trip and on-trip TI. The results indicate benefits ranging from 12.7 to 15.5 SEK per passenger, i.e., almost double the reference values in the literature, which is possibly due to methodological differences and/or case-specific factors. The system perspective illustrates and compares the ATIS system benefits with the total direct costs of delays per weekday. The results indicate such systems could have reduced the average delay costs by 23%. This suggests that implementing ATIS has the potential for cost savings in mitigating the impact of disruptions. Such savings are even more substantial when including transfers, e.g., commuter networks with multiple lines.

## 5.2 Future works

As mentioned earlier, the case study is based on a single line of a commuter rail system which can overlook potential network effects and benefits, e.g., connections, from interactions between different lines and transportation networks. The presentation of the valuation of TI can be further disaggregated in terms of travel periods (e.g., peak, or off-peak), and share of different types of benefits (e.g., waiting time, travel time uncertainty). Moreover, the generalization to other transportation systems, e.g., long-distance trains, may require adjustments to consider the specificities of such systems, e.g., other cost parameters, infrequent departures, and unavailable ridership data.

Future research could address some of the limitations that were mentioned in the introduction. For instance, a broader case study including multiple lines and/or modes of transport could offer a more general valuation. Another direction for future works is to account for the system dynamics and variability in the event of disruptions by considering equilibrium states instead of static modelling using the passenger perspective. Such work can be used to explore optimal diversion strategies for passengers using traffic information to seek alternative (undisrupted) routes.

## Declarations
**Funding**
This research is part of a research project funded by a grant from the Swedish Transport Administration (Trafikverket).

**Acknowledgements**
The authors are grateful to Disa Asplund, Hanna Lindgren and Ulrica Sörman for discussions and improvement recommendations. We are also grateful to Martin Joborn for earlier comments and suggestions, and to Per Lingvall for project administration.

**Conflict of interests**
The authors have no conflicts of interest to declare that are relevant to the content of this article.

**Ethical approval**
This research did not involve any studies with human participants or animals performed by any of the authors.

## References

Ahlberg, J. 2015. Kostnader för störningar i infrastrukturen: metodik och fallstudier på väg och järnväg. *VTI notat.* Linköping: Statens väg- och transportforskningsinstitut.

Ait-Ali, A. & Eliasson, J. 2019. Dynamic Origin-Destination Estimation Using Smart Card Data: An Entropy Maximisation Approach. *Preprint arXiv:1909.02826*.

Ait-Ali, A. & Eliasson, J. 2021. European Railway Deregulation: An overview of market organization and capacity allocation. *Transportmetrica A: Transport Science*, 1-27.

Ait-Ali, A. & Peterson, A. 2023. Assessing the effects of traffic information to passengers during disruptions: a literature review. *World Conference on Transport Research - WCTR 2023.* Montreal: Forthcoming in Transportation Research Procedia.

Al-Deek, H. & Kanafani, A. 1993. Modeling the benefits of advanced traveler information systems in corridors with incidents. *Transportation Research Part C: Emerging Technologies,* 1**,** 303-324.

Arentze, T. A. & Timmermans, H. J. P. 2005. Information gain, novelty seeking and travel: a model of dynamic activity-travel behavior under conditions of uncertainty. *Transportation Research Part A: Policy and Practice,* 39**,** 125-145.

Asplund, D. 2021. *Optimal frequency of public transport in a small city: examination of a simple method*, VTI.

Ben-Elia, E. & Avineri, E. 2015. Response to travel information: A behavioural review. *Transport reviews,* 35**,** 352-377.

Berggren, U., Brundell-Freij, K., Svensson, H. & Wretstrand, A. 2021. Effects from usage of pre-trip information and passenger scheduling strategies on waiting times in public transport: an empirical survey based on a dedicated smartphone application. *Public Transport,* 13**,** 503-531.

Brakewood, C., Barbeau, S. & Watkins, K. 2014. An experiment evaluating the impacts of real-time transit information on bus riders in Tampa, Florida. *Transportation Research Part A: Policy and Practice,* 69**,** 409-422.

Brakewood, C. & Watkins, K. 2019. A literature review of the passenger benefits of real-time transit information. *Transport Reviews,* 39**,** 327-356.

Bruglieri, M., Bruschi, F., Colorni, A., Luè, A., Nocerino, R. & Rana, V. 2015. A Real-time Information System for Public Transport in Case of Delays and Service Disruptions. *Transportation Research Procedia,* 10**,** 493-502.

Börjesson, M. & Eliasson, J. 2011. On the use of "average delay" as a measure of train reliability. *Transportation Research Part A: Policy and Practice,* 45**,** 171-184.
Cats, O. & Gkioulou, Z. 2017. Modeling the impacts of public transport reliability and travel information on passengers' waiting-time uncertainty. *EURO Journal on Transportation and Logistics,* 6**,** 247-270.
Cats, O., Koutsopoulos, H. N., Burghout, W. & Toledo, T. 2011. Effect of Real-Time Transit Information on Dynamic Path Choice of Passengers. *Transportation Research Record,* 2217**,** 46-54.
Caulfield, B. & Mahony, M. O. 2007. An Examination of the Public Transport Information Requirements of Users. *IEEE Transactions on Intelligent Transportation Systems,* 8**,** 21-30.
Chorus, C. G. 2012. Travel Information: Time to Drop the Labels? *IEEE Transactions on Intelligent Transportation Systems,* 13**,** 1235-1242.
Chorus, C. G., Arentze, T. A., Molin, E. J. E., Timmermans, H. J. P. & Van Wee, B. 2006. The value of travel information: Decision strategy-specific conceptualizations and numerical examples. *Transportation Research Part B: Methodological,* 40**,** 504-519.
Dziekan, K. & Kottenhoff, K. 2007. Dynamic at-stop real-time information displays for public transport: effects on customers. *Transportation Research Part A: Policy and Practice,* 41**,** 489-501.
Dziekan, K. & Vermeulen, A. 2006. Psychological Effects of and Design Preferences for Real-Time Information Displays. *Journal of Public Transportation,* 9**,** 1-19.
Eltved, M., Breyer, N., Ingvardson, J. B. & Nielsen, O. A. 2021. Impacts of long-term service disruptions on passenger travel behaviour: A smart card analysis from the Greater Copenhagen area. *Transportation Research Part C: Emerging Technologies,* 131**,** 103198.
Fearnley, N., Leiren, M. D., Skollerud, K. H. & Aarhaug, J. Benefit of measures for universal design in public transport. European Transport Conference, 2009Association for European Transport (AET), 2009.
Ferris, B., Watkins, K. & Borning, A. 2010. OneBusAway: results from providing real-time arrival information for public transit. *Proceedings of the SIGCHI Conference on Human Factors in Computing Systems.* Atlanta, Georgia, USA: Association for Computing Machinery.
Frohne, E., Kildor, Tomiwoj & Schröter, U. 2014. Linjekarta för Stockholms pendeltåg. Wikimedia Commons.
Ghahramani, N. & Brakewood, C. 2016. Trends in mobile transit information utilization: An exploratory analysis of transit app in New York City. *Journal of Public Transportation,* 19**,** 9.
Giannopoulos, G. A. 2004. The application of information and communication technologies in transport. *European Journal of Operational Research,* 152**,** 302-320.
Halpern, N. 2021. The provision of service information for public transport.
Kattan, L., de Barros, A. G. & Saleemi, H. 2013. Travel behavior changes and responses to advanced traveler information in prolonged and large-scale network disruptions: A case study of west LRT line construction in the city of Calgary. *Transportation Research Part F: Traffic Psychology and Behaviour,* 21**,** 90-102.
Kenworthy, J. R. 2020. Sustainable mobility in ten Swedish cities: a comparative international assessment of urban transport indicators in Stockholm, Göteborg, Malmö, Linköping, Helsingborg, Uppsala, Örebro, Västerås, Jönköping, Umeå and Freiburg im Breisgau, Germany. *K2 working paper:*.
Lappin, J. E. 2000. Advanced Traveler Information Service (ATIS): Who Are ATIS Customers? *In:* JOHN, A. V. (ed.) *National Transportation Systems Center.*
Lehtonen, M. & Kulmala, R. 2002. Benefits of Pilot Implementation of Public Transport

Signal Priorities and Real-Time Passenger Information. *Transportation Research Record,* 1799**,** 18-25.

Leng, N. & Corman, F. 2020. The role of information availability to passengers in public transport disruptions: An agent-based simulation approach. *Transportation Research Part A: Policy and Practice,* 133**,** 214-236.

Lu, R. 2015. The effects of information and communication technologies on accessibility.

Malchow, M., Kanafani, A. & Varaiya, P. 1996. The economics of traffic information: a state-of-the-art report.

Müller, S. A., Leich, G. & Nagel, K. 2020. The effect of unexpected disruptions and information times on public transport passengers: a simulation study. *Procedia Computer Science,* 170**,** 745-750.

Paulsen, M., Rasmussen, T. K. & Nielsen, O. A. 2021. Impacts of real-time information levels in public transport: A large-scale case study using an adaptive passenger path choice model. *Transportation Research Part A: Policy and Practice,* 148**,** 155-182.

Peer, S. 2013. *The economics of trip scheduling, travel time variability and traffic information*.

Pirra, M., Diana, M. & Castro, A. 2017. Information provision in public transport: Indicators and benchmarking across europe. *Transport Infrastructure and Systems.* CRC Press.

Prather Persson, C. 1998. *The Railway Station and The Interregional Traveller-traveller preferences and implications for the planning process.* Lund University.

Pritchard, J. Providing Improved Crowding Information to Provide Benefits for Rail Passengers and Operators. *In:* STANTON, N. A., ed. Advances in Human Aspects of Transportation, 2018 2018 Cham. Springer International Publishing, 973-984.

Rich, J. & Vandet, C. A. 2019. Is the value of travel time savings increasing? Analysis throughout a financial crisis. *Transportation research part A: policy and practice,* 124**,** 145-168.

Skoglund, T. 2012. *Investigating the Impacts of ICT-Mediated Services–the Case of Public Transport Traveller Information*, Chalmers Tekniska Hogskola (Sweden).

Tang, L., Ho, C. Q., Hensher, D. & Zhang, X. 2020. Demand for travel information: what, when and how much is required by urban travellers.

Teng, J., Wang, H. & Lei, X. 2018. Analyzing the Effects of Public Transit Information Systems Based on Mobile Terminal on Choice Behavior of Travelers on Route. *CICTP 2017.*

Titov, W., Gerlach, C. & Schlegel, T. 2023. Information Provision on Interactive Smart Public Displays in Public Transport in Events of Disruptions. *In:* GESA PRAETORIUS, SELLBERG, C. & PATRIARCA, R. (eds.) *AHFE (2023) International Conference. .* USA: AHFE Open Access,. AHFE International.

Trafikverket 2020. Analysmetod och samhällsekonomiska kalkylvärden för transportsektorn: ASEK 7.0.

Trafikverket 2023. Trafikinformation – Resenärsnöjdhet inom järnväg : Årsrapport 2022. *Trafikverkets publikationer.* Borlänge.

Tseng, Y.-Y., Knockaert, J. & Verhoef, E. T. 2013. A revealed-preference study of behavioural impacts of real-time traffic information. *Transportation Research Part C: Emerging Technologies,* 30**,** 196-209.

van Essen, M., Thomas, T., van Berkum, E. & Chorus, C. 2016. From user equilibrium to system optimum: a literature review on the role of travel information, bounded rationality and non-selfish behaviour at the network and individual levels. *Transport Reviews,* 36**,** 527-548.

Waara, N. Older and disabled people's need and valuation of traveller information in public transport.  Proceeding of The Association for European Transport Conference,

2009. 1-21.

Wardman, M. 2001. A review of British evidence on time and service quality valuations. *Transportation Research Part E: Logistics and Transportation Review,* 37**,** 107-128.

Wardman, M., Hine, J. & Stradling, S. 2001. Interchange and Travel Choice, Vol. 1, Vol. 2 and Research Summary. *Edinburgh: Central Research Unit, Scottish Executive*.

Watkins, K. E., Ferris, B., Borning, A., Rutherford, G. S. & Layton, D. 2011. Where Is My Bus? Impact of mobile real-time information on the perceived and actual wait time of transit riders. *Transportation Research Part A: Policy and Practice,* 45**,** 839-848.

Wu, W., Juan, Z. & Luo, Q. 2010. Short-term Choice of Travelers with Effect of Traffic Information. *Journal of Transportation Systems Engineering and Information Technology,* 10**,** 100-105.

Yeboah, G., Cottrill, C. D., Nelson, J. D., Corsar, D., Markovic, M. & Edwards, P. 2019. Understanding factors influencing public transport passengers' pre-travel information-seeking behaviour. *Public Transport,* 11**,** 135-158.

Zito, P., Amato, G., Amoroso, S. & Berrittella, M. 2011. The effect of Advanced Traveller Information Systems on public transport demand and its uncertainty. *Transportmetrica,* 7**,** 31-43.